\documentclass[aps,prb,twocolumn,superscriptaddress,showpacs]{revtex4-1}
\usepackage{mathbbol,amsmath,amsfonts,amssymb,bm,dsfont}
\usepackage{color,graphicx,psfrag}
\usepackage{epstopdf}

\def\im{{\sf i}}
\def\Z{{\mathds{Z}}}

\begin{document}

\title{Changing anyonic ground degeneracy with engineered gauge fields}

\author{Emilio Cobanera}
\affiliation{Institute for Theoretical Physics, Center for Extreme Matter and Emergent Phenomena, Utrecht University, Leuvenlaan 4, 3584 CE Utrecht, The Netherlands}
\email[Electronic address: ]{Coban003@uu.nl}

\author{Jascha Ulrich}
\affiliation{JARA-Institute for Quantum Information, RWTH Aachen University,
	52074 Aachen, Germany}

\author{Fabian Hassler}
\affiliation{JARA-Institute for Quantum Information, RWTH Aachen University,
	52074 Aachen, Germany}

\date{\today}

\begin{abstract}
For systems of lattice anyons like Majorana and parafermions, the
unconventional quantum statistics determines a set of global symmetries (e.g.,
fermion parity for Majoranas) admitting no relevant perturbations.  Any
operator that breaks these symmetries explicitly would violate locality if
added to the the Hamiltonian.  As a consequence, the associated
quasi-degeneracy of topologically non-trivial phases is protected, at least
partially, by locality via the symmetries singled out by quantum statistics.
We show that it is possible to bypass this type of protection by way of
specifically engineered gauge fields, in order to modify the topological
structure of the edge of the system without destroying the topological order
completely.  To illustrate our ideas in a concrete setting, we focus on the
\(\Z_6\) parafermion chain.  Starting in the topological phase of the chain
(sixfold ground degeneracy), we show that a gauge field with restricted
dynamics acts as a relevant perturbation, driving a transition to a phase with
threefold degeneracy and \(\Z_3\) parafermion edge modes.  The transition from
the \(\Z_3\) to the topologically trivial phase occurs on a critical line in
the three-state Potts universality class.  Hence, to all effects and purposes,
the gauged \(\Z_6\) chain realizes the \(\Z_3\) parafermion chain.  We also
investigate numerically the emergence of Majorana edge modes when the \(\Z_6\)
chain is coupled to a differently restricted gauge field.  
%
\end{abstract}

\maketitle

\section{Introduction}

The pursuit of topological quantum information processing with 
mesoscopic arrays\cite{terhal2012, nussinov2012, vijai:15} of Majorana
chains\cite{kitaev2001} is grounded on well investigated solid state 
technologies and showcases a new bridge between condensed matter 
physics and engineering. With current experimental techniques, naturally 
available topological quantum matter is ill suited for quantum information 
processing.\cite{willett2013} Hence, one should try to design synthetic 
topological quantum matter with enhanced desirable characteristics. 
Mesoscopic physics naturally sets the stage for this proposal because 
the microscopic understanding of interacting topological quantum matter 
is incomplete.  Hence one would like to exploit naturally occurring topological 
states, not as they come but as building blocks in mesoscopic arrays.

The Majorana chain itself for example is not naturally available, 
and so the proposals for its realization\cite{alicea12,beenakker13} 
are good examples of the quest for synthetic topological quantum matter. 
The parafermion chain\cite{fendley2012} is an even more ambitious proposal
along these lines,\cite{alicea15} motivated by a problem that parafermions
cannot, at least directly,\cite{mong2014} wholly resolve in the end: the 
limited computational power of Majoranas. With parafermions the 
idea\cite{clarke2013, lindner2012} is to modify mesoscopically the edge 
of a fractional quantum spin Hall insulator by placing alternating 
\emph{s}-wave superconducting and ferromagnetic islands in order to 
obtain, at low energies, an effective 1d lattice system of hybridized
parafermion zero-energy modes. This basic mesoscopic blueprint has
excited theoretical interest in various types of parafermion chains, 
with varying degrees of exoticism in their quantum phase 
diagrams\cite{vaezi13,milsted2014,zhuang15,li15}  
and also varying potential for mesoscopic realization.\cite{milsted2014}

Parafermions, and in particular Majoranas, are examples of topological
zero-modes. Such modes are good for quantum information processing
for three basic reasons: they are local, isolated in energy, and stable.  
As a consequence, the manifold of ground states of the system that hosts 
these modes inherits desirable properties. First, its degeneracy is stable.
Second, it is possible to perform rotations in the ground manifold
by coupling local probes to the topological modes. And third,
there is an energy gap to non-topological excitations that sets the 
time scale for these manipulations.  

In spite of their crucial practical value, the stability of topological
zero-energy modes poses a problem for incremental design.  Take for example
the basic blueprint for parafermions of the previous paragraph.  The quantum
statistics of the emergent lattice parafermion field is characterized by an
integer \(p\geq 2\), and on the mesoscopic setups of
Refs.~\onlinecite{clarke2013,lindner2012} for example, \(p\) can only take the
values \(p=2m\), with \(m\) an odd, positive integer.  As a consequence,
topological qutrits (corresponding to \(p=3\)) are not included in this
otherwise very attractive proposal.  This is a problem for the circuit model
assisted by topological information processing, since quantum algorithms with
qutrits are the best studied ones after qubits.  Now, since the ground
degeneracy of the \(p=6\) parafermion chain is six, it is natural to seek for
a design that modifies the basic platform minimally, just enough to halve the
degeneracy.  But the degeneracy is protected by the stability of the
topological zero-energy modes!  It should be clear from this example how, in
general, incremental design principles can be at odds with ``topological''
ones.

In this paper, we will propose a method based on symmetry manipulation without
symmetry breaking to modify in a controlled fashion the structure of the set
of zero-energy modes at the edge of topological quantum matter.  Our method in
a nutshell starts by identifying discrete gauge fields naturally present
in mesoscopic setups and naturally coupled to protecting symmetries. After 
providing dynamics to the gauge field, some or all of the topological 
degeneracy disappears. However, if the gauge fluctuations are properly 
manipulated, it is possible to remove only some of the topological degeneracy. 
As a consequence, the topological structure of the edge 
changes in a controlled and predictable way: some zero-modes disappear and 
the remaining ones control the new ground manifold of the system.

Given the symmetry analysis underlying our idea, one may ask why not just
break explicitly the symmetry in question.  An explicit symmetry breaking term
would in general obtain the same result as that obtained by the gauge field
without breaking the symmetry, and, in one space dimension, these two
scenarios are often connected by a duality transformation.  The answer is
rooted deeply in the structure of anyon fields.  As we will show, due to anyon
statistics, the symmetries that must be manipulated in our framework cannot be
explicitly broken without violating locality.  Effectively, this observation
provides a symmetry-based explanation of the stability, or ``protection," of
the zero-energy modes.  By coupling the symmetry to a gauge field rather than
breaking it, we manage to modify topological zero modes without breaking
locality.  This idea evolved from our investigation of the effect of phase
slips on the Majorana chain,\cite{vanHeck2014} since they obtain a \(\Z_2\)
gauge field minimally coupled to the Majorana degrees of freedom.

The outline of the paper is as follows.  In Sec.~\ref{sec:kitaev_model}, we
explain our general ideas using a familiar example, the Majorana chain of
Kitaev.  The key point is that a \(\Z_2\) gauge field arise naturally in 
physical realizations of topological chains in terms of charged fermions.  
Making this local $\mathds{Z}_2$ gauge field dynamic by mesoscopic
manipulation allows to modify the topological edge structure without explicitly 
breaking fermion-parity. In the following sections, we generalize these ideas 
to the parafermion chain. In Sec.~\ref{pfs}, we recall the basic theory 
of parafermions, discuss the parafermionic generalization of the Majorana 
chain and show that the global symmetry protecting the parafermionic zero-modes 
cannot be broken explicitly without violating locality. In Sec.~\ref{gauged_pfs}, 
we introduce the gauged parafermion chain where parafermions of degree $p$ 
are naturally coupled to a $\mathds{Z}_p$ gauge field. By means of  
an exact duality transformation we show that by restricting the dynamics 
of the \(\Z_p\) gauge field to a nontrivial subgroup of $\mathds{Z}_p$ 
we can obtain parafermions of reduced degree without explicitly breaking
the protecting symmetry. We show evidence that our construction works as 
predicted by investigating the period of the supercurrent in a gauged 
parafermion ring junction as a function of the strength of the gauge fluctuations.
Restricting the gauge field dynamics to a $\mathds{Z}_2$ subgroup, we obtain a
supercurrent periodicity consistent with $\mathds{Z}_3$ parafermionic edge
modes.  In Section\,\ref{q_phase_diag}, we confirm in another way that we 
have $\mathds{Z}_3$ parafermionic zero modes in this gauged \(\Z_6\) chain
by investigating the phase diagram of the system with the DRMG
algorithm. The critical exponents of the transition between the 
topologically non-trivial and trivial phase correspond to the
universality class of the three-state Potts model for any finite strength of
restricted gauge fluctuations, that is, on a critical line. We also 
compute the phase diagram of the model with effective $\mathds{Z}_2$
parafermionic zero modes (Majoranas). In both cases, the modified
topological zero modes are stable and exist in an extended region 
of coupling space.  We conclude with a summary and outlook in Sec.~\ref{outlook}.

\section{Destroying the edge modes of the Kitaev model with gauge fields} 
\label{sec:kitaev_model}

In this section we briefly recall the gauged Majorana 
chain\cite{cobanera_holo,vanHeck2014} in order to keep a concrete 
example in mind as we proceed to investigate the more complex case 
of parafermions. 

Let us introduce two Hermitian Majorana operators $a_j$, $b_j$
per site site $j=1,\dots,L$ of a chain, such that
\begin{eqnarray}
\{a_i, b_j\} = 0,\quad 
\{a_i, a_j\} = \{b_i, b_j\} = 2 \delta_{ij}.\nonumber
\end{eqnarray}
Majoranas are related to Pauli matrices, 
\begin{eqnarray}\label{JW}
a_i=\sigma^x_i\prod_{j=1}^{i-1}\sigma^z_j,\quad
b_i=\sigma^y_i\prod_{j=1}^{i-1}\sigma^z_j,
\end{eqnarray}
and fermionic creation and annihilation operators,
\begin{eqnarray}
a_i&=&c_i+c_i^\dagger,\label{majofermi1}\\ 
-\im b_i&=&-c_i+c_i^\dagger=c_i(-1)^{n_i}+c_i^\dagger.\label{majofermi2}
\end{eqnarray}
Let \(\hat{F}= \sum_{i=1}^L n_i\) denote the total fermion number operator. 
Then the operator of fermionic parity
\begin{eqnarray}
(-1)^{\hat{F}}=\prod_{i=1}^L(-\im a_ib_i)
\end{eqnarray}
measures the total number of fermions modulo two. 

The Majorana chain of Kitaev \cite{kitaev2001} is specified by the Hamiltonian 
\begin{align}\label{eq:kitaev_model}
H_{\sf k} =-\im h\sum_{i=1}^L a_ib_i - \im J \sum_{j=1}^{L-1} b_{j} a_{j+1},
\end{align}
describing the competition between a topologically non-trivial phase for
\(|h/J|<1\) and a trivial phase for the complementary regime. The points
\(|h/J|=1\) are critical. In its topologically non-trivial phase, the Majorana
chain displays twofold quasi-degeneracy protected by fermionic parity
$(-1)^{\hat{F}}$. This symmetry cannot be broken explicitly due to a superselection 
rule enforcing vanishing matrix elements between sectors distinguished by 
fermionic parity. Such a superselection rule can be traced back to the 
requirement of locality of the physical Hamiltonian. 

It has been shown in Ref.~\onlinecite{bravyi:10} that the 
quasi-degeneracy of the topologically non-trivial phase cannot be lifted 
by local perturbations. 
This result suggests that there is no way of modifying the topological 
structure of the edge by local perturbations. But this inference is in 
fact incorrect for a physical realization, as opposed to the ideal 
representation, of the Majorana chain.\cite{vanHeck2014} The reason is 
that the Majorana chain
will typically be realized in terms of electrically charged fermions, e.g.,
electrons, which carry the electromagnetic $U(1)$ gauge structure. 
Induced superconductivity breaks most of that symmetry, but a $\mathds{Z}_2$ 
part survives.\cite{reznik89} This residual gauge symmetry, 
corresponding to an invariance of energy levels under a local 
multiplication of the Majorana operators $a_j$, $b_j$ by a factor of 
$-1$, is no longer manifest in the Majorana chain due to a tacit but 
specific gauge choice.  Restoring the gauge fields normally 
used to absorb the effects of these residual gauge transformations leads 
to the Hamiltonian
\begin{align}\label{eq:gauged_kitaev_model}
H_{\sf gk}=-\im h\sum_{i=1}^L a_ib_i
-\sum_{j=1}^{L-1}\bigl[\im J b_{j} a_{j+1} \tau_j^z+\kappa\tau_j^x\bigr].
\end{align}
of the gauged Majorana chain\cite{cobanera_holo,vanHeck2014}
The \(\Z_2\) gauge field is represented by Pauli matrices $\tau_j^x$, 
$\tau_j^z$, placed on the links $j=1,\dots,L-1$ of the chain. 
For $\kappa \neq 0$, the gauge field is quantum dynamic.\cite{footnote2} 

The gauged Majorana chain must be supplemented with gauge constraints
identifying physical states.  Since the $U(1)$ symmetry originates from
electromagnetism, all physical states should obey the $\mathds{Z}_2$
descendant of Gauss law.  The chain has local symmetries
\begin{align} \label{eq:kitaev_gauge_constraints}
G_1 &= (-\im a_1 b_1) \tau_1^x,\quad G_L=\tau^x_{L-1}(-\im a_Lb_L),\\
G_j &= \tau_{j-1}^x (-\im a_j b_j) \tau_j^x \quad (j=2,\dots,L-1), 
\end{align}
and enforcing the Gauss law corresponds to demanding that the local symmetries
$G_j$ for $j = 2, \dots, L$ of the gauged Majorana chain act as the identity
on physical states, $G_j = \mathds{1}$.\cite{fradkin}

The perturbation proportional to \(h\) in Eq.~\eqref{eq:gauged_kitaev_model} 
plays a subtle but definite role in tying the fate of the topologically 
non-trivial phase to the gauge field. The reason may be traced back to 
the effect of the gauge field on the topological boundary modes of the 
Majorana chain. For \(h=0\), the gauged Majorana chain displays exact 
zero-energy modes
\begin{eqnarray}\label{eq:kitaev_zero_modes}
[H_{\sf gk}|_{h=0},a_1]=0=[H_{\sf gk}|_{h=0},b_L],
\end{eqnarray}
perfectly localized at the end points of the chain and completely insensitive 
to the gauge field. This insensitivity disappears as soon as \(h\neq0\). 
In the absence of gauge fluctuations, \(\kappa=0\), it is still possible 
to compute explicitly the evolution of the edge modes with \(h\). We obtain  
\begin{eqnarray}
\gamma_l&=&
a_1+\sum_{i=2}^L\Big(\frac{h}{J}\Big)^{i-1}a_i\tau^x_{i-1}\dots\tau_1^x,\\
\gamma_r&=&
b_L+\sum_{i=1}^{L-1}\Big(\frac{h}{J}\Big)^{L-i} b_i\tau^x_i\dots \tau^x_{L-1},
\end{eqnarray}
with
\begin{eqnarray}
{}[H_{\sf gk}|_{\kappa=0},\gamma_l]&=&
2\im J\Big(\frac{h}{J}\Big)^Lb_L\tau^x_{L-1}\dots\tau^x_1,\quad\\
{}[H_{\sf gk}|_{\kappa=0},\gamma_r]&=&
-2\im J\Big(\frac{h}{J}\Big)^L a_1\tau^x_1\dots \tau^x_{L-1}.
\end{eqnarray}
The gauge field and Majorana (quasi) zero-energy boundary modes are 
now explicitly intertwined. In general, any gauge invariant perturbation, 
no matter how small, will suffice to intertwine the topological zero modes 
with the gauge field.  

For some physical setups, due to the stiffness of the phase of the
superconducting condensate, the dynamics of the gauge field is strongly
suppressed, $\kappa \approx 0$.  In this case, the gauge fields may be removed
from the theory by a specific gauge choice in order to recover the
topologically non-trivial Majorana chain~\eqref{eq:kitaev_model}.  However,
we have showed in a previous publication\cite{vanHeck2014} that a finite value
for $\kappa$ may be obtained through quantum phase slips in a mesoscopic
implementation of the Kitaev model.  At finite $\kappa$, the properties of the
Kitaev model are fundamentally altered, because $\kappa$ measures the strength
of a relevant perturbation of fermionic parity.  Any non-zero value of
$\kappa$ will remove the topological degeneracy of the Majorana chain for
large enough system size.  But why is this the case?

One explanation comes from solving the gauge
constraints \eqref{eq:kitaev_gauge_constraints} for $\tau_j^x$,
giving
\begin{eqnarray}\label{solvez2}
\tau_j^x = \prod_{k=1}^{j} (-\im a_k b_k) = 
(-1)^{\hat{F}} \prod_{k=j+1}^{L} (-\im a_kb_k).
\end{eqnarray}
In this way the dynamics of the gauge field $\tau_j^z$ drives nonlocal
parity measurement of the part of the chain to the left and to the 
right of site $j$, which is, in general, incompatible with the topological
degeneracy. While the gauge fields are local degrees of freedom, they 
have highly nonlocal effects on the electronic degrees of freedom, which 
makes them prime candidates for the manipulation of topological phases
without breaking protecting symmetries. Moreover, the effects we
are describing are not tied to the mean field approximation. 
Very recently there has been some progress in the investigation
of particle number conserving Majorana chains,\cite{ortiz14,iemini15}
and, as should be clear from our arguments, we would not expect our 
conclusions to be any different for the \(\Z_2\) gauged versions of 
such models.

To summarize, discrete gauge theories can emerge in a natural, 
controllable way in physical realizations of topological zero-energy 
boundary modes. Moreover, for the Majorana chain, one can destroy the 
topologically non-trivial phase phase by gauging the global
symmetry of fermionic parity and making the corresponding $\mathds{Z}_2$ 
gauge field dynamic. For parafermions, a richer structure is possible.
This is the subject of the rest of the paper.

\section{Protection of the edge modes of the parafermion chain by locality}
\label{pfs} 

In this section we will briefly recall the basics of parafermions in one
dimension, introduce the parafermion chain as a natural generalization of the
Majorana chain, and explain how the ground degeneracy of the parafermion chain
is protected by the interplay between locality and anyonic statistics. There is
a very recent review on parafermions Ref.~\onlinecite{alicea15} that the reader
may consult for further details on implementation proposals.  

Parafermions are described in terms of a natural generalization
of the Majorana operators of the previous section. A one-dimensional, 
lattice-regularized parafermion field, defined on sites
$i=1, \dots, L$, has the following properties \cite{alcaraz1981}
\begin{eqnarray}
\Gamma_i\Delta_j&=&\omega\Delta_j\Gamma_i\quad (i\leq j),
\label{exch1}\\
\Gamma_i\Gamma_j&=&\omega\Gamma_j\Gamma_i\quad (i<j),
\label{exch2} \\ 
\Delta_i\Delta_j&=&\omega\Delta_j\Delta_i \quad (i<j),
\label{exch3}\\
\Gamma_i^{p}=&\mathds{1}&=(-1)^{p-1}\Delta_i^{p},\label{exc}
\end{eqnarray}
where $\omega = e^{i 2\pi/p}$ is a $p$-th root of unity.  The order
\(p=2,3,\dots\) of the parafermion field characterizes the exchange
(Eqs.~\eqref{exch1}, \eqref{exch2}, and \eqref{exch3}) and exclusion
(Eq.~\eqref{exc}) statistics of these degrees of freedom.\cite{cobanerafp} In
addition, the parafermion field is unitary,
\begin{eqnarray}
\Gamma_i\Gamma_i^\dagger=\mathds{1}=\Delta_i\Delta_i^\dagger.\label{unitary}
\end{eqnarray}

Up to a phase, parafermions of order $p = 2$ are Majorana operators.
More precisely, 
\begin{eqnarray}\label{pfsp2}
\Gamma_j = a_j, \quad \Delta_j=-\im b_j \qquad \quad (p=2).
\end{eqnarray} 
It is in general true of everything that follows that setting \(p=2\) will
recover some standard facts about fermions. 

Following the reasoning for the construction of the Majorana 
chain by pairing the second parafermion on each site with strength 
$J$ with the first one the next site, one obtains the parafermion 
chain\cite{fendley2012}
\begin{eqnarray}\label{pf_chain}
H_{\sf p}=-\frac{h}{2}\sum_{i=1}^L(\Gamma_i^\dagger \Delta_i+H.c.)-
\frac{J}{2}\sum_{i=1}^{L-1}(\Delta_i\Gamma_{i+1}^\dagger+H.c.).\ \ \ \ 
\end{eqnarray}
For \(h=0\), the parafermion chain displays exact zero-energy 
parafermion edge modes,
\begin{eqnarray}\label{edge_modes}
[H_{\sf p},\Gamma_1]=0=[H_{\sf p},\Delta_L],
\end{eqnarray} 
perfectly localized on the end points of the chain. As a consequence,
the ground energy level is $p$-fold degenerate, and, in fact, every energy
level is at least \(p\)-fold degenerate. This is the topological degeneracy
of the parafermion chain.  Small values of \(h\)\cite{footnote3} will not remove the 
topological (quasi-)degeneracy of the ground energy level, but may remove the
topological degeneracy of excited energy levels.\cite{fendley2012,jermyn14}

For the Majorana chain there is a strong link between topological
degeneracy and fermionic parity. Is there something similar for parafermions? 
In order to obtain a precise answer to this question it is necessary to relate the 
parafermion field to a charged  field and an underlying Fock space. Let us 
introduce creation and annihilation operators \(C_i^\dagger,C_i\) satisfying 
the relations\cite{cobanerafp}
\begin{eqnarray}
C_i^{\dagger\, p}&=&0=C_i^{p},\\
C_i^lC_i^{\dagger\, l}&+&C_i^{\dagger\, (p-l)}C_i^{p-l}=\mathds{1},\\
C_iC_j&=& \omega C_jC_i\quad (i<j),\\
C_iC_j^\dagger&=& \bar \omega C_j^\dagger C_i\quad (i<j),
\end{eqnarray}
(\(l=1,\dots,p-1\)) 
where the bar denotes complex conjugation, $\bar \omega = e^{- i 2\pi/p}$.
The number operator for these ``Fock parafermions'' \cite{cobanerafp} is
\begin{equation}\label{countpfs}
N_i=\sum_{l=1}^{2m-1}C_i^{\dagger\, l}C_i^l,
\end{equation}
and the Fock vacuum satisfies \(C_i|0\rangle=0,\ \forall i\). 
Let \(\hat{P}= \sum_{i=1}^L N_i\) denote the total parafermion number 
operator. Then 
\begin{eqnarray}
\omega^{\hat{P}}=e^{\im 2\pi \hat{P}/p}
\end{eqnarray}
measures the parafermionic parity.

The analogy between  \(\omega^{\hat{P}}\) and \((-1)^F\)  
is compelling, but it is not clear yet whether \(\omega^{\hat{P}}\)
is a symmetry of the parafermion chain. The relation between Fock 
and standard parafermions is\cite{cobanerafp}
\begin{eqnarray}
\Gamma_i&=&C_i+C_i^{\dagger\, (p-1)},\label{parafermi1}\\
\Delta_i&=&C_i\omega^N_i+C_i^{\dagger\, (p-1)}.\label{parafermi2}
\end{eqnarray}
It is illuminating to compare Eqs. \eqref{majofermi1} and 
\eqref{parafermi1}, and Eqs. \eqref{majofermi2} and \eqref{parafermi2}.
Since
\(C_i^{\dagger\,(p-1)}\omega^{N_i}=C_i^{\dagger\,(p-1)}\), then
\begin{eqnarray}
\Gamma_i^\dagger\Delta_i=
C_i^\dagger C_i\omega^{N_i}+C_i^{p-1}C_i^{\dagger\,(p-1)}=\omega^{N_i}.
\end{eqnarray}
Hence,
\begin{eqnarray}
\omega^{\hat{P}}=\prod_{i=1}^L\Gamma_i^\dagger \Delta_i.
\end{eqnarray}
It is straightforward now to check that the parafermionic parity is indeed a
symmetry of the parafermion chain.   

Similar to the fermionic parity $(-1)^{\hat{F}}$ in the Majorana case, the
parafermionic parity $\omega^{\hat{P}}$ distinguishes the various ground states of the parafermion
chain.  The parafermion field transforms under this symmetry as
\begin{eqnarray}\label{dp}
\bar{\omega}^{\hat{P}} \Gamma_i\omega^{\hat{P}}=\omega \Gamma_i,\quad 
\bar{\omega}^{\hat{P}} \Delta_i \omega^{\hat{P}}=\omega \Delta_i.
\end{eqnarray}
The stability of the parafermionic zero modes is explained by the fact that
$\omega^{\hat{P}}$ is always a symmetry of any {\textit local} Hamiltonian of
parafermions: for these systems, the global \(\Z_p\) symmetry
\(\omega^{\hat{P}}\) cannot be explicitly broken without violating locality.

Why is this so? In order to answer this question, it is useful to introduce 
an explicit realization of parafermions analogous to 
the relation between Majoranas and Pauli matrices, Eq.~\eqref{JW}. 
First, associate to each site of the chain a pair of unitary 
\(p\times p\) matrices \(U_i,V_i\), often called clock or circulant 
matrices, that commute on different sites and otherwise satisfy 
\begin{eqnarray}
\label{clocks}
V_iU_i=\omega U_iV_i,\quad
V_i^p=U_i^p=\mathds{1}.  
\end{eqnarray} 
We can represent a single pair of clock matrices as
\begin{eqnarray}\label{clock_matrices} \hspace*{-0.5cm}
{U}=\begin{pmatrix}
1& 0& 0& \cdots& 0\\
0&\omega & 0& \cdots& 0\\
0& 0&\omega^2 & \cdots& 0\\
\vdots& \vdots& \vdots&      & \vdots \\
0& 0& 0& \cdots& 0\\
0& 0& 0& \cdots&\omega^{p-1} \\
\end{pmatrix} ,\quad
{V}=
\begin{pmatrix}
0& 1& 0& \cdots& 0\\
0& 0& 1& \cdots& 0\\
0& 0& 0& \cdots& 0\\
\vdots& \vdots& \vdots&      & \vdots \\
0& 0& 0& \cdots& 1\\
1& 0& 0& \cdots& 0\\
\end{pmatrix},  
\label{VsandUs}
\end{eqnarray}
Then,
\begin{eqnarray}\label{pfmJW}
\Gamma_i=V_iU_{i-1}\dots U_1,\quad \Delta_i=\Gamma_iU_i.  
\end{eqnarray} 
In this representation we see explicitly the string responsible for the
non-trivial exchange statistics. 
Now it is possible to show that operators \(O_i\) localized in a finite 
region around site \(i\), meaning that they are constructible as a product 
of parafermions in that region, commute for disjoint regions if and only if 
they are invariant under the symmetry \(\omega^{\hat{N}}\) of modular
conservation of parafermion number.  As a consequence any local Hamiltonian 
of parafermions is invariant under this symmetry as well: this symmetry cannot 
be explicitly broken without breaking locality. 

Let us point out in closing an interesting relation between Fock parafermions
and electron fractionalization. For \(p=2m\) with \(m\) odd, the Fock parafermion 
field is literally\cite{cobanerafrp} the \(m\)th rood of the electron field. 
More explicitly, the composite fields \(C_i^m,C_i^{\dagger\, m}\) satisfy 
canonical anticommutation relations,
\begin{eqnarray}
\{C_i^m,C_j^m\}=0,\quad \{C_i^m,C_j^{\dagger\, m}\}=\delta_{i,j}.
\end{eqnarray}
Tunneling between electronic and parafermionic systems can then be modeled by
terms of the form \(-\Gamma f^\dagger C^m+H.c.\), where \(f\) is a canonical 
fermionic annihilation operator. Practical applications of this observation
were explored in Ref.~\onlinecite{cobanerafrp}. 

For the mesoscopic setups of Refs.~\onlinecite{clarke2013,lindner2012}, it is
indeed the case that the integer \(m\) characterizes the fractionalization in
the system, since the quasiparticles of the fractional spin Hall insulator in
the mesoscopic array carry charge \(e/m\).  The superconductors on the edge
induce pairing, and the relation \(p=2m\) is then a reflection of the fact
that it takes \(2m\) quasiparticles to form a Cooper pair.  It is interesting
to notice that the \(m=1\) is allowed.  In this simplest case, there is no
fractionalization.  The 2d insulator is just a spin Hall insulator, and the
parafermion field becomes (up to a phase) a lattice Majorana field.
Ref.~\onlinecite{alicea15} offers an up to date overview of proposals for
realizing parafermions in mesoscopic arrays.

\section{Changing the edge modes of the parafermion chain with gauge fields}
\label{gauged_pfs}

In this section we will show how to selectively remove some of the zero-energy
edge modes of the parafermion chain.  By analogy to the gauged Majorana chain,
the basic idea is to first restore the gauge symmetry of the parafermion
chain, and then allow for suitably engineered gauge fluctuations in order to
split in energy some gauge sectors and not others.  The sectors that remain
unsplit support parafermionic topological edge modes of reduced order.


The generalization of the $\mathds{Z}_2$ gauge field of the Majorana chain 
is the $\mathds{Z}_p$ gauge field $\tau_i, \sigma_i$ satisfying
\begin{eqnarray}\label{gauge_fields}
\tau_i^{p}=\sigma_i^{p}=\mathds{1},\quad \tau_i\sigma_i=\omega\sigma_i\tau_i
\end{eqnarray} 
and commuting otherwise. This algebra is the same algebra as for
the clock matrices \(U_i,V_i\) of the previous section, but we use 
a different notation here for clarity. Then the Hamiltonian of the 
gauged parafermion chain is
\begin{eqnarray}\label{gaugedpfs}
H_{\sf gp}&=&-\frac{h}{2}\sum_{i=1}^L(\Gamma_i^\dagger \Delta_i+H.c.)\\
&-&\sum_{i=1}^{L-1}\Big[\frac{J}{2}(\Delta_i\Gamma_{i+1}^\dagger\tau_i+H.c.)
+\kappa(\sigma_i^a+H.c.)\Big].
\nonumber
\end{eqnarray}
In general, the design of the mesoscopic setup meant to realize
the parafermion chain provides control over the gauge fluctuations, 
characterized by \(\kappa\) and an integer \(a=0,\dots,p-1\). 
The values \(\kappa=0\) and/or \(a=0\) correspond to quenched gauge 
fluctuations, that is, no fluctuations at all. For \(a>1\) we say that
the gauge fluctuations are restricted. The reason will become clear
as we proceed.

The local \(\Z_p\) symmetries of the chain are 
\begin{eqnarray}
G_1=(\Gamma_1^\dagger\Delta_1)\sigma_1^\dagger,\quad 
G_L=\sigma_{L-1} (\Gamma_L^\dagger\Delta_L),\\
G_i=\sigma_{i-1}(\Gamma_i^\dagger\Delta_i)\sigma_i^\dagger\quad (i=2,\dots,L-1),
\end{eqnarray}
They seem a priori independent of the details of the gauge fluctuations, and,
being local, they cannot be spontaneously broken.\cite{elitzur75} What is then
the meaning of the of the integer parameter \(a\)?  The answer is that local
gauge fluctuations characterized by \(a>1\) generate highly nonlocal
measurements of parafermion number modulo $p/a$ which split the ground state
multiplet, leaving behind a topological degeneracy of reduced degree.  This can
be seen by removing explicitly the gauge redundancy and performing a duality
transformation, as we will show next. Later in this section we will show that
this reduced ground state degeneracy corresponds to  topological parafermionic
edge modes of reduced degree.

\subsection{Changing topological degeneracy with restricted gauge
fluctuations}

Let us start by rewriting the gauged parafermion chain in 
terms of the clock matrices of Eq.~\eqref{clock_matrices}, 
\begin{eqnarray}\label{pert_gclock}
H_{\sf gp} =&-&
\frac{h}{2}\sum_{i=1}^{L}(U_i +U_i^\dagger) \\
&-&\sum_{i=1}^{L-1}\Big[\frac{J}{2}(V_iV_{i+1}^\dagger\tau_i+H.c.)+
\frac{\kappa}{2}(\sigma_i^a+\sigma_i^{\dagger\,a})\big],\nonumber
\end{eqnarray}
by virtue of the transformation Eq.~\eqref{JW}.  The resulting model is akin
to, but not quite the same as a \(\Z_p\) Higgs model if \(a\neq1\).  In terms
of clock matrices, the gauge symmetries are $G_1 = U_1\sigma_1^\dagger$, $G_i
= \sigma_{i-1}U_i \sigma_i^\dagger$ for $i=2,\dots,L-1$ and $G_N =
\sigma_{N-1} U_{N}$. The global symmetry $\omega^{\hat{P}}$ simply reads 
\begin{align}\label{parafermionic_parity_clock}
\omega^{\hat{P}} = \prod_{i=1}^L U_i. 
\end{align}

By introducing simultaneous eigenstates $|\{u_j\}\rangle$ of the 
clock matrices $U_j$,
\begin{eqnarray}
U_j |\{u_j\}\rangle = \omega^{u_j} |\{u_j\}\rangle\quad (j=1,\dots,L),
\end{eqnarray}
and simultaneous eigenstates $|\{\sigma_j\}\rangle$ of the clock matrices
$\sigma_j$,
\begin{eqnarray}
\sigma_j |\{\sigma_j\}\rangle = \omega^{\sigma_j} |\{\sigma_j\}\rangle
\quad (j=1,\dots,L-1)
\end{eqnarray}
we can define gauge-fixed states on the tensor-product space of the matter 
and gauge fields. They are of the form
\begin{align}
|\{u_j\}\rangle_{\text{gf}} 
= |\{u_j\}\rangle \otimes | \{\sigma_j = \sum_{i> j} (\alpha_i-u_i)\} \rangle,
\end{align}
with integers $\alpha_i$ defined for $i=2,\dots,L$.  The states
belonging to the gauge sector labeled by the $\alpha_i$  obey 
the relation
\begin{align}
G_i |\{u_j\}\rangle_{\text{gf}} &= \omega^{\alpha_i}
|\{u_j\}\rangle_{\text{gf}} \quad (i = 2,\dots,L), 
\end{align}
and the physical sector corresponds to the choice $\alpha_i = 0$.  Projecting
the Hamiltonian onto the gauge-fixed states leads to the gauge-fixed
Hamiltonian
\begin{eqnarray}\label{eq:parafermion_chain}
H_{\sf{gp}}^{\text{gf}}=&-&\frac{h}{2} \sum_{i=1}^L (U_i + U_i^\dag) \\
&-&\sum_{i=1}^{L-1}\Big[\frac{J}{2} V_i V_{i+1}^\dag 
+ \frac{\kappa}{2} \prod_{j > i} (\omega^{a \alpha_j} U_j^{\dagger a}) + H.c.\Big],
\nonumber
\end{eqnarray}
where the operator $U_i$ acts in the natural way on the gauge-fixed states,
\begin{eqnarray}
U_j |\{u_j\}\rangle_\text{gf} = \omega^{u_j} |\{u_j\}\rangle_\text{gf}.
\end{eqnarray}  

For $\kappa = 0$, the gauge-fixed parafermion chain
\(H_{\sf{gp}}^{\text{gf}}\) is just the celebrated clock model (see
Ref.~\onlinecite{ortiz:12} for an up-to-date review), and one could translate
back to parafermions to recover the standard parafermion chain.  But for
\(\kappa\neq 0\), the gauge-fixed Hamiltonian shows explicitly how local gauge
fluctuations at finite $\kappa$ affect the parafermionic degrees of freedom in
a highly nonlocal way.  The strings in \(H_{\sf{gp}}^{\text{gf}}\) are not
local neither in terms of clock or parafermion degrees of freedom. On the
parafermions, they correspond to nonlocal measurements of parafermion number
modulo $p/a$. 

To see why this corresponds to a reduction of the ground state degeneracy, we
note that the gauge-fixed Hamiltonian may be brought into a local form 
by means of the duality transformation
\begin{align}
V_i &\mapsto \prod_{j=1}^{i} U_j \quad (i=1,\dots,L),\nonumber \\
U_i &\mapsto V_i^\dag V_{i+1} \quad (i = 1,\dots,L-1),\\ 
U_L &\mapsto V_L^\dag, \nonumber
\end{align}
exact for finite chains.\cite{ortiz:12} The resulting Hamiltonian is
\begin{align}\label{dual_ham}
H_{\sf gp}^{\text{gf}, D} =& -\frac{h}{2}(V_L^\dagger+H.c.) -
\frac{h}{2}\sum_{i=1}^{L-1}(V_i^\dagger V_{i+1}+H.c.)\\
&-\sum_{i=1}^{L-1}\Big[\frac{J}{2}U_{i+1}^\dagger+
\frac{\kappa}{2} V_{i+1}^a\,\prod_{j>i}\omega^{a \alpha_j} + H.c.\Big]
\nonumber
\end{align}
and the global symmetry $\omega^{\hat{P}}$ is mapped onto a boundary symmetry,
$\omega^{\hat P} \mapsto V_1^\dagger$.

The Hamiltonian \(H_{\sf gp}^{\text{gf}, D}\) makes the role of the gauge
fluctuations particularly clear.  Let us focus on the physical sector
$(\alpha_i = 0$) for simplicity.  Then, if we neglect finite-size corrections,
that is, the boundary term \(hV_L/2+H.c.\), \(H_{\sf gp}^{\text{gf}, D}\)
takes the simpler form $\sum_i H_i$, with
\begin{eqnarray}
H_i = -\Big[\frac{J}{2}U_{i+1}^\dagger+
\frac{\kappa}{2}V_{i+1}^a+\frac{h}{2}V_i^\dagger V_{i+1}\Big] + H.c.
\end{eqnarray}
%
%
Let \(a/p=a'/r\), with \(a'\) and \(r\) relative primes, and suppose for
simplicity that \(p=rs\) with \(r\) and \(s\) also relative primes.  Then, the
equivalent $\omega^{\hat{P}_D}$ of Eq.~\eqref{parafermionic_parity_clock} with
the matrices $U_i$ of the dual model \eqref{dual_ham}, gives
\begin{eqnarray}
\bar{\omega}^{r \hat{P}_D}V_i^a \omega^{r\hat{P}_D}=V_i^a,
\end{eqnarray}
since \(e^{\im 2\pi ar/p}=1\). Hence, the ``longitudinal field" 
\(\kappa\) breaks the symmetry \(\omega^{\hat{P}_D}\), but not 
completely if \(a>1\): The \(\Z_s\) symmetry \(\omega^{r\hat{P}_D}\)
survives the gauge fluctuations unscathed. Conversely, the 
\(\Z_r\) symmetry \(\omega^{s\hat{P}_D}\) is completely
broken. In this way, gauge fluctuations obtain specific 
relevant perturbations that reduce but do not remove
completely the topological degeneracy of the parafermion
chain. 

\subsection{Effect of restricted gauge fluctuation on 
topological boundary modes}

\begin{figure*}
\includegraphics[angle=0,width=1.\textwidth]{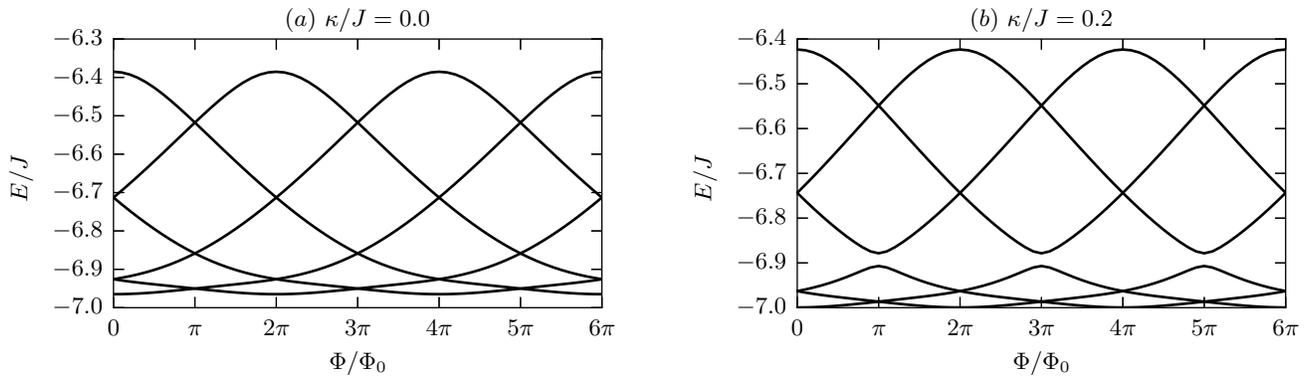}
\caption{Lowest energy levels of the Hamiltonian \eqref{gaugedpfs_ring}
modeling a $\mathds{Z}_6$ parafermionic ring chain as a function of the
enclosed flux $\Phi$.  The $\mathds{Z}_6$ parafermions are coupled to
$\mathds{Z}_6$ gauge fields whose dynamics is restricted to $\mathds{Z}_2$.
($a$) For static gauge fields with $\kappa = 0$, the spectrum shows the
$12\pi$ flux periodicity expected for $\mathds{Z}_6$ parafermions.  ($b$) For
dynamic gauge fields with \(\kappa=0.2 J\), the flux periodicity is reduced
from $12 \pi$ to a flux periodicity of $6 \pi$ consistent with effective
$\mathds{Z}_3$ parafermions.} \label{fig1}
\end{figure*}

Restricted gauge fluctuations obtain a bulk relevant perturbation that
partially removes the topological degeneracy of the parafermion chain.  What
is the effect on the edge modes?  Here we address this question in terms of
the Josephson effect, generalizing the celebrated anomalous periodicity of the
supercurrent in the Majorana chain for investigating gauged Parafermion
chains.

Let us take \(p>a=s>1\) for  concreteness, so that \(a/p=1/r\), 
with \(r\) and \(s\) relative primes. Then the exact zero modes 
\begin{eqnarray}\label{left_behind}
\tilde{\Gamma}_1=\Gamma_1^r,\quad \tilde{\Delta}_L=\Delta_L^r
\end{eqnarray}
of the gauged Parafermion chain with \(h=0\) realize parafermions
of reduced degree \(s\). These topological modes commute
with the symmetry \(\omega^{s\hat{P}}\) that is effectively,
but not explicitly, broken by the restricted gauge fluctuations.
Hence, we expect these zero modes to survive the gauge fluctuations 
because they live in sectors of constant \(\omega^{s\hat{P}}\) 
and will not become split in energy.

We can make this picture quantitative and very close to the 
situation for the gauged Majorana chain if we recall a technical
result about clock matrices. Let us write \(\tau_{i}^{(q)},\sigma^{(q)}_i\)
for the clock matrices of order \(q=r,s\). Then we have the
identifications\cite{cobanerafp} 
\begin{eqnarray}
\sigma_i\cong \sigma^{(r)}_i\otimes\sigma^{(s)}_{i},\quad 
\tau_i\cong(\tau^{(r)}_{i})^n\otimes(\tau^{(s)}_{i})^n, 
\end{eqnarray}
with \(n\) the unique solution of\cite{footnote4} 
\begin{eqnarray}
(r+s)n=1\ \ (\mbox{mod}\ p).
\end{eqnarray}
In particular,
\begin{eqnarray}
\sigma_i^s\cong (\sigma^{(r)}_{i})^s\otimes\mathds{1}^{(s)}_{i}.
\end{eqnarray}
Since\cite{footnote5}
\begin{eqnarray}
 (\tau^{(r)}_{i})^n(\sigma^{(r)}_{i})^s=e^{\im 2\pi /r}
(\tau^{(r)}_{i})^n (\sigma^{(r)}_{i})^s,
\end{eqnarray}
we can make the simpler identification
\begin{eqnarray}\label{useful}
\sigma_i^s\cong\sigma^{(r)}_{i}\otimes\mathds{1}^{(s)}_{i} ,\quad 
\tau_i\cong\tau^{(r)}_i\otimes\tau^{(s)}_i.
\end{eqnarray}

In terms of the identification of Eq.~\eqref{useful}, we can 
rewrite the gauged Parafermion chain in reduced form, 
\begin{eqnarray}\label{gaugedpfs_reduced}
H_{\sf gp}'=&-&\frac{h}{2}\sum_{i=1}^L(\Gamma_i^\dagger \Delta_i+H.c.)\\
&-&\sum_{i=1}^{L-1}[\frac{J}{2}(\alpha_i\Delta_i\Gamma_{i+1}^\dagger\tau^{(r)}_i +H.c.)
+\kappa(\sigma^{(r)}_i+H.c.)]. \nonumber
\end{eqnarray}
The site-dependent numerical phases \(\alpha_i\) are \(s\) roots of unity
obtained from diagonalizing the \(\tau^{(s)}_{i}\). Since, in particular,
\(\alpha_i^p=1\), they can be removed by a gauge choice in the definition 
of the parafermion field. Hence the situation is as follows: A parafermion 
chain coupled to a \(\Z_p\) gauge field with restricted fluctuations 
(\(a=s=p/r\)) is equivalent to a parafermion chain coupled to a \(\Z_r\) 
gauge field with unrestricted gauge fluctuations (\(a=1\)). Based on our 
experience with the gauged Majorana chain, we expect that the \(\Z_r\) 
topological boundary modes
\begin{eqnarray}
\tilde{\tilde{\Delta}}_1=\Delta_1^s,\quad \tilde{\tilde{\Delta}}_L=\Delta_L^s
\end{eqnarray}
will be gapped out by the \(\Z_r\) gauge field, leaving behind the 
\(\Z_s\) zero modes of Eq.~\eqref{left_behind}.

We can check this claim explicitly by extending the theory of the 
\(4\pi\) Josephson effect in the Majorana chain to our gauged parafermion 
chain, or more precisely, to \(\tilde{H}_{\sf pg}\) to avoid 
unnecessary numerical cost. To drive a supercurrent across the gauged 
parafermion chain it is convenient to close the system into a ring 
threaded by a flux \(\Phi\). The Hamiltonian for this parafermion ring 
junction is 
\begin{eqnarray}\label{gaugedpfs_ring}
H'(\Phi) = H_{\sf gp}'
-\frac{J}{2}(C_L^\dagger \tau^{(r)}_{L} C_1 e^{-\im 2\Phi/p}+H.c.)
\end{eqnarray}
(the Fock parafermions \(C_i\), \(C_i^\dagger\) of charge \(2e/p\) were 
introduced in the previous section). The tunneling term is also coupled 
to the gauge field in order to preserve gauge invariance. The gauge 
symmetries of the ring junction are
\begin{eqnarray}\label{ring_gauge_syms}
G_i'=\sigma^{(r)}_{i-1} (\Gamma_i^\dagger\Delta_i) (\sigma^{(r)}_{i})^\dagger
\quad (i=i+L).
\end{eqnarray}

Let us consider for concreteness a \(\Z_2\) gauge field coupled to
parafermions of order \(p=6\), so that \(r=2\) and \(s=3\). 
Fig.~\ref{fig1}($a$) shows the six lowest energy levels of 
\(H_{\sf gp}(\Phi)\) in the \(\Z_2\) gauge invariant sector and 
for \(\kappa=0=h\). Any gauge sector would show the same result since 
they are all degenerate in energy. The full \(12\pi\) Josephson effect 
expected of the usual \(p=6\) parafermion chain\cite{clarke2013,cobanerafp} 
is clearly visible. For \(\kappa/J=0.2\) and \(h=0\), the period of the 
supercurrent has decreased to \(6\pi\), c.f.~Fig.~\ref{fig1}($b$). The 
zero modes responsible for this periodicity are precisely the parafermionic 
zero modes \(\tilde{\Gamma}_1,\tilde{\Delta}_L\) of order \(s=3\).  

We will show in Sec.~\ref{q_phase_diag} that the term proportional
to \(\kappa\) is a relevant perturbation. As the system size increases, 
smaller and smaller values of \(\kappa\) suffice to produce an appreciable 
change in the periodicity of the supercurrent.

\section{Stability against local perturbations at zero temperature}
\label{q_phase_diag} 

As we saw in the previous section, gauge fluctuations change the 
topological edge structure of the parafermion chain. What is 
the stability of the modified edge against a generic perturbation? 
In order to gain some insight into this problem, in this section we 
determine numerically the phase diagram of the gauged parafermion chain
 in the coupling space defined by \(h\) and \(\kappa\). Recall that the 
perturbation of strength \(h\) favors a trivial pairing of parafermions. 
For numerical investigations it is convenient to explicitly 
remove the gauge redundancy of the gauged parafermion chain. Hence in this 
section we work exclusively with the dual Hamiltonian 
\(H_{\sf gp}^{\text{gf}, D}\) of Eq.~\eqref{dual_ham} in the gauge
invariant sector. 

Since energy levels, and quantum phase diagrams in particular, are 
insensitive to the distinction between parafermion and clock chains, 
let us emphasize once more the conceptual issue at stake. The gauged 
parafermion chain and the clock Hamiltonian of Eq.~\eqref{pert_gclock},
or its gauge-fixed version, are the same Hamiltonian from an spectral
point of view. However, if one were to realize \(H_{\sf gp}\) on a physical 
platform, then the representation Eq.~\eqref{gaugedpfs} would be naturally 
suited for a fermionic platform\cite{clarke2013,lindner2012}, and
that of Eq.~\eqref{pert_gclock} a bosonic one, e.g., bosonic cold atoms.  
However, for the Hamiltonian \(H_{\sf gp}\) written as in Eq.~\eqref{pert_gclock}, 
it is easy to break some or all of its global symmetries explicitly with 
perturbations of the form \(-h_s\sum_{i=1}^L (V_i^s+H.c.)\). No gauge 
field is needed to effectively generate such terms. Hence, in a cold 
atom realization, the ground degeneracy of \(H_{\sf gp}\) is fragile.\cite{nussinov09}  
In contrast, such perturbations are non-local in terms of parafermions 
and strictly absent in fermionic realizations, except, effectively, 
as gauge fields!

In the following, we will focus on parafermions of order \(p=6\).  Then it is
possible to restrict the gauge field dynamics to $\mathds{Z}_r$ with $r = 2,
3$ by taking \(a=3,2\) respectively.  We call the corresponding models the 
\(\Z(6,2)\) and \(\Z(6,3)\) chains.

\subsection{Critical exponents for $\mathds{Z}(6, 2)$} 

In the previous section, we have discussed that increasing $h$ should
eventually drive the system from a topological phase into a trivial phase.  If
we indeed have $\mathds{Z}_3$ parafermionic zero modes for finite $\kappa$,
as suggested by the supercurrent numerical calculation of Sec.~\ref{gauged_pfs},
this must be reflected in the universality class of the transition.  To
confirm the presence of $\mathds{Z}_3$ parafermionic zero modes, we calculate
the critical exponents of the transition by means of a detailed finite-size
scaling analysis.

For the analysis of the system, we use the Hamiltonian\,\eqref{dual_ham} in
the physical sector $\alpha_i = 0$ which decomposes into subspaces $H_q$ to
eigenvalue $e^{i 2\pi q/6}$ of the conserved quantity $V_1$.  We assume that a
single relevant operator associated with $h$ will drive a transition from a
symmetry-broken $\mathds{Z}_3$ phase into a trivial phase.  Consequently, a
scaling analysis along the lines of the Ising transition in absence of
symmetry-breaking fields (in the bulk) applies.\cite{cardy} While the bulk
terms of $H_q$ are invariant under the $\mathds{Z}_3$ symmetry, the operator
$v_N$ and the operator $(e^{-i 2\pi q/6} v_2 + H.c)$ arising from the
projection on $H_q$ explicitly breaks this symmetry at the boundary.  Although
negligible in the thermodynamic limit, those symmetry-breaking boundary
operators are expected to give rise to irrelevant scaling fields leading to
increased finite-size effects.  \cite{campostrini:14}

For the numerical determination of the location of the critical point and the
critical exponents, we use the density renormalization group (DMRG) algorithm
and variationally determine the eigenstates of the Hamiltonian
Eq.~\eqref{dual_ham} in the ground state sector $q=0$, requiring
for each variational state $|\psi\rangle$ with (approximate) eigenvalue
$\lambda$ a precision of at least $\sqrt{\langle \psi | (H-\lambda I)^2 | \psi
\rangle}/\lambda \leq 10^{-7}$.  We analyze systems with system size $N=16,
32, 64, 128$ sites and maximal internal bond dimensions $D=128$.  

\begin{figure}[tbp] \centering
\includegraphics[width=1.\columnwidth]{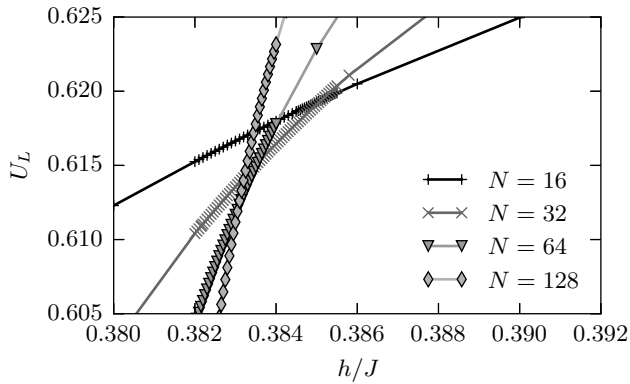} 
\caption{Behavior of the Binder cumulants $U_L=1-\langle m^4 \rangle/3
  \langle m^2\rangle^2$, defined in terms of the magnetization $m = \sum_{i=1}^N
  (v_i + v_i^\dag)/2N$, for systems of different sizes $N$ along the line
  $h=\kappa$ in vicinity of the crossing points.  \label{fig:z6_binder}}
\end{figure}

Let us start with a detailed finite-size scaling analysis along the line $h =
\kappa$.  The similarity to the Ising case suggests exploiting the Binder
cumulant $U_L = 1 - \langle m^4 \rangle/3 \langle m^2 \rangle^2$ of the
``magnetization per site'' $m = \sum_{i=1}^N (v_i + v_i^\dag)/2N$ for a precise
determination of the location of the critical point.  Considering one
irrelevant scaling field with renormalization group eigenvalue $-\omega < 0$
and expanding the scaling fields to lowest non-vanishing order, the Binder
cumulant satisfies the scaling relation\cite{pelissetto:02} 
\begin{align}\label{eq:binder_scaling}
U_L = \Phi\bigl( |h - h_c| N^{1/\nu}, u_0 N^{-\omega} \bigr).
\end{align}
Here $\Phi$ is a universal scaling function, $\nu$ is the critical 
exponent associated with the divergence of the correlation length, 
$h_c$ is the critical point, and $u_0$ is the value of the irrelevant 
scaling field at the critical point. As before, $N$ denotes the system 
size.  By expanding $\Phi$ 
around the critical point $(0, u_0)$, we find that the crossing points
$h_{N_1,N_2}$ of the Binder cumulants of different systems with size $N_1$,
$N_2$ scale as
\begin{align}
|h_{N_1,N_2} - h_c| \propto N_1^{-1/\nu - \omega} \frac{1 - (N_2/N_1)^{-\omega}}{1 -
  (N_2/N_1)^{-1/\nu}}
\end{align}
and thus converge to the critical point $h_c$ as $N_1$ goes to infinity for
fixed ratios $N_1/N_2$.

Figure Fig.~\ref{fig:z6_binder} shows the behavior of the Binder cumulants for
different system sizes $N=16, 32, 64, 128$ and $D=128$.  We obtain the
location of the critical point and the exponent $-1/\nu - \omega$ using a
Shanks transformation for the data tuples $(N_1,N_2) = (16,32), (32, 64),
(64,128)$.  Since we only have 3 data points available, we estimate the errors
by varying $h_c$ and observing when the data points $|h_{N_1,N_2} - h_c|$ in a
double-logarithmic plot deviate visibly from a line, yielding $1/\nu + \omega
\approx 1.69$ and $h_c = 0.38426(5)$.  In absence of irrelevant scaling
fields, the scaling form Eq.~\eqref{eq:binder_scaling} predicts that the
values of $U_L$ of different system sizes $N$ collapse onto a single curve as
a function of $z = (h-h_c) N^{1/\nu}$.  The failure of a data collapse
assuming the absence of irrelevant scaling fields, using $1/\nu \approx 1.69$,
confirms the non-negligible role of irrelevant scaling fields for the scaling
analysis.

\begin{figure}[tbp] \centering
\includegraphics[width=1.\columnwidth]{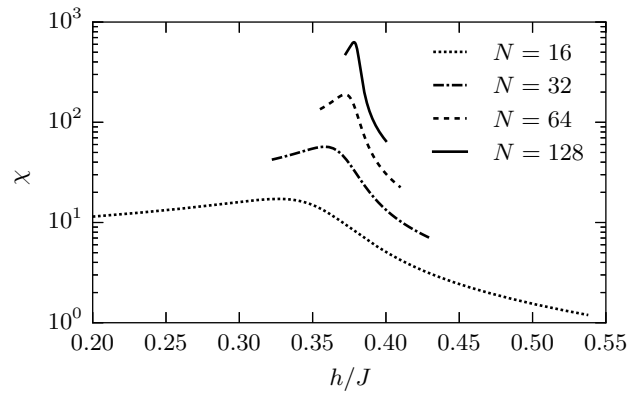} 
\caption{Behavior of the susceptibility $\chi$ 
  along the line $h=\kappa$ for different system sizes $N$. Data points are
  not indicated for better visual clarity.\label{fig:z6_suscep}}
\end{figure}
\begin{figure}[tbp] \centering
\includegraphics[width=1.\columnwidth]{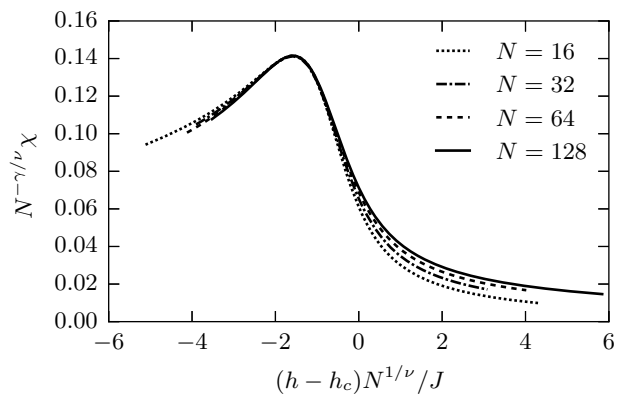} 
\caption{Data collapse of the susceptibility data shown in
  Fig.~\ref{fig:z6_suscep} with $\nu = 0.832$, $\gamma/\nu = 1.731$ determined
  from a minimization of the residuals
  Eq.~\eqref{eq:suscep_collapse_residuals} in an interval of size $\Delta z =
  \pm 0.4$ around the susceptibility maximum.  The critical point $h_c$ is
  fixed at the value obtained using Binder cumulants.  Away from the maximum,
  no complete data collapse is obtained.  To the right of the maximum, the
  curves for different $N$ approach the maximum $N_\text{max}=128$ curve
  monotonously from below, whereas to the left of the maximum, they approach
  the $N_\text{max}$ curve monotonously from above.  This systematic structure
  of the deviations suggests contributions due to irrelevant scaling fields,
  possibly associated with the symmetry-breaking boundary operators in the
  Hamiltonian $H_q$, as a possible explanation.  Data points are not indicated
  for better visual clarity, the data point spacing is roughly $\Delta z =
  0.1$ for all curves.
\label{fig:z6_suscep_collapse}}
\end{figure}

For the determination of the critical exponents $\nu$, $\gamma/\nu$, we
consider the susceptibility 
\begin{eqnarray}
\chi = N^2(\langle m^2 \rangle - \langle m
\rangle^2).
\end{eqnarray}
In absence of irrelevant scaling fields, \(\chi\) satisfies the scaling relation
\begin{align}\label{eq:suscep_scaling}
  \chi = N^{\gamma/\nu} \tilde \Phi\bigl( |h-h_c| N^{1/\nu}\bigr),
\end{align}
where $\tilde \Phi$ is a universal scaling function and $\gamma$ is the
critical exponent associated with the divergence of the susceptibility at the
critical point.  The behavior of the susceptibility across the transition is
shown in Fig.~\ref{fig:z6_suscep} for different system sizes.  According to
the scaling form Eq.~\eqref{eq:suscep_scaling}, the susceptibility values
$\chi N^{-\gamma/\nu}$ obtained for different system sizes collapse onto a
single curve as a function of $z = (h-h_c)N^{1/\nu}$.  
Then, for given values of $z$ and $h_c$ obtained using Binder cumulants, we 
extract the critical exponents $\nu$, $\gamma/\nu$ from the susceptibility 
values by minimizing the quadratic differences 
\begin{align}\label{eq:suscep_collapse_residuals}
  \sum_z \sum_{i < j} \Bigl[ \chi_i(h_c + z N_i^{-1/\nu})
  N_i^{-\gamma/\nu}\nonumber \\
  - \chi_j(h_c + z N_j^{-1/\nu}) N_j^{-\gamma/\nu} \Bigr]^2
\end{align}
\begin{figure}[tbp] \centering
\includegraphics[width=1.\columnwidth]{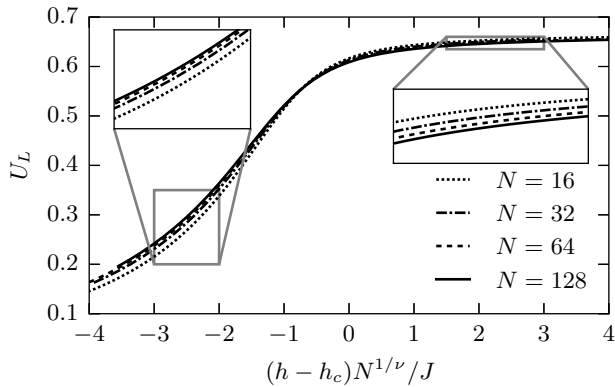} 
\caption{Data collapse of the Binder cumulants $U_L$ obtained using $h_c$,
  $\nu$ determined from Binder cumulants and susceptibility.  Data points are
  not indicated for better visual clarity, the data point spacing is roughly
  $\Delta z = 0.1$ for all curves.  Insets show close ups of the curves in the
  marked regions, showing that the distance to the $N_\text{max} = 128$ curve
  systematically decreases as $N$ is increased.
  \label{fig:z6_binder_collapse}}
\end{figure}%
\begin{figure}[tbp] \centering
\includegraphics[width=1.\columnwidth]{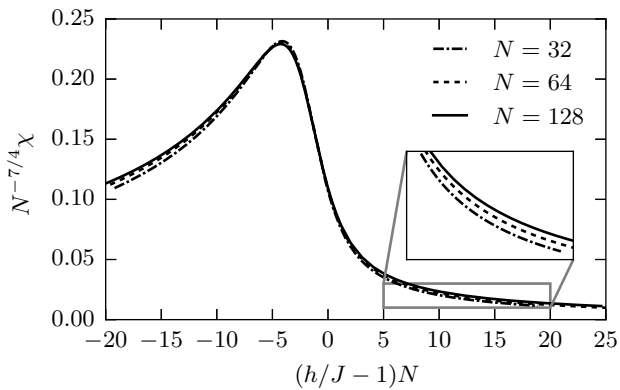} 
\caption{Data collapse of the susceptibility curves obtained for the model
  \eqref{dual_ham} along the line $\kappa = 0$ with $p=2$ which
  corresponds to an Ising model with symmetry-breaking boundary conditions,
  using the exactly known critical point $h_c/J = 1$ and the critical
  exponents $\nu = 1$, $\gamma = 7/4$.  The collapse shows a structure similar
  to the collapse of the susceptibility data obtained along the line $\kappa =
  h$ for $p=6$ that was shown in Fig.~\ref{fig:z6_suscep_collapse}, with
  deviations from perfect data collapse away from the susceptibility
  peak.}\label{fig:ising_suscep_collapse_exact}
\end{figure}%
between the rescaled susceptibility curves $\chi_i(h)$ of systems of size
$N_i$.  Performing the fit in a range of $\Delta z/J = \pm 0.4$ around the
maximum of the scaled susceptibility, which is the most prominent feature, we
obtain the data collapse shown in Fig.~\ref{fig:z6_suscep_collapse} with
$\gamma/\nu = 1.731(2)$, $\nu = 0.832(3)$.  

The error estimates are obtained by checking when data collapses obtained 
using values $\nu + \Delta \nu$,
$\gamma/\nu + \Delta (\gamma/\nu)$ for the critical exponents, with $\nu$,
$\gamma/\nu$ as quoted above and $\Delta \nu$, $\Delta (\gamma/\nu)$ some
perturbation, show appreciable deviations from the data collapse of
Fig.~\ref{fig:z6_suscep_collapse}.  While by choice of the fit region, the
collapse shown in Fig.~\ref{fig:z6_suscep_collapse} is very good around the
maximum, it shows systematic deviations away from it.  We note in particular
that the distance to the maximal size $N_\text{max}=128$ curve is decreasing
monotonously as $N$ is increased.  Performing the fit around other regions of
the susceptibility curve yields data collapses with deviations that do not
show a similar internal consistency.  A data collapse of the Binder cumulant
$Q_L$ obtained with $h_c$ and $\nu$ as determined above shows reasonable data
collapse with deviations showing the same systematic behavior, see
Fig.~\ref{fig:z6_binder_collapse}.

The systematic behavior of the deviations suggests irrelevant scaling fields
due to the symmetry-breaking boundary operators in the Hamiltonian $H_q$ as a
plausible explanation.  In order to show that this explanation is indeed
consistent, we have simulated the model \eqref{dual_ham} with $p=2$ along the
line $\kappa = 0$, where it is just an Ising model with symmetry-breaking
boundary conditions.  The collapse of the susceptibility curves of systems
with different sizes $N_i$, using the exactly known values for the critical
point $h_c/J = 1$ and the critical exponents $\nu = 1$, $\gamma = 7/4$, is
shown in Fig.~\ref{fig:ising_suscep_collapse_exact}.  The data shows good
collapse around the susceptibility maximum and deviations away from it, i.e.,
a structure similar to the collapse discussed previously.  This shows that
symmetry-breaking boundary operators may indeed be the cause of the irrelevant
scaling fields and strongly supports the assumption that our choice of fit
region around the susceptibility maximum allows us to extract the critical
exponents with a negligible systematic error.

Summing up, we find $\nu = 0.832(3)$, $\gamma/\nu = 1.731(2)$ for the critical
exponents of the transition along the line $h = \kappa$. This result is in
good agreement with the exact values $\nu = 5/6 = 0.8\bar 3$, $\gamma/\nu =
26/15 = 1.7\bar 3$ expected for the three-state Potts model.  \cite{wu:82}
We thus numerically confirm the transition from a $\mathds{Z}_3$ to a trivial
phase that was suggested by the structure of the dual Hamiltonian\,\eqref{dual_ham}
and the periodicity of the supercurrent discussed in Sec.\,\ref{gauged_pfs}.

\begin{figure}[tbp] \centering
\includegraphics[width=1.\columnwidth]{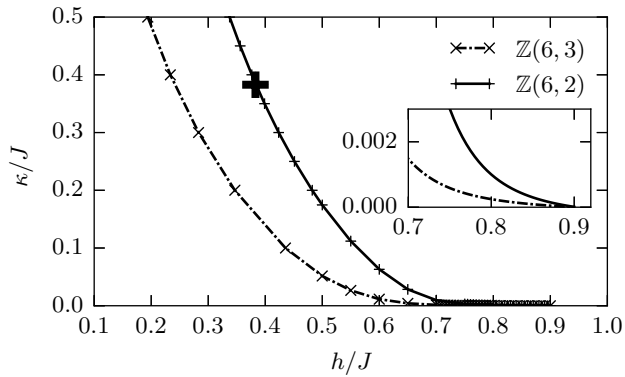}
\caption{Location of the critical line for the transition from the nontrivial
  phase (to the left of the line) to the trivial phase (to the right of the
  line) for the $\mathds{Z}(6, 2)$ and $\mathds{Z}(6, 3)$ model.  The critical
  line is determined from the crossing points of the Binder cumulants $U_L$
  for system sizes $(N_1, N_2) = (32, 64)$ and the Binder cumulants are
  calculated numerically using the DMRG algorithm with internal bond dimension
  $D=64$.  Data points are marked as small crosses and linearly interpolated
  for visual guidance.  The large cross indicates the position of the point
  where we have determined the critical exponents $\nu = 0.832$, $\gamma/\nu =
  1.731$ consistent with a $\mathds{Z}_3$ phase\cite{wu:82} by means of a more
  detailed finite-size scaling analysis.  The inset shows a closeup of the
  region around $\kappa = 0$, showing that in both cases, the critical lines
  of the models end at the same point at $\kappa = 0$.
  }\label{fig:phase_diagram}
\end{figure}

\subsection{Phase diagram}

According to our findings for the $\mathds{Z}(6, 2)$ model, the crossings of
the Binder cumulants indicate the location of the critical point up to small
finite-size corrections.  In fact, this is expected to hold regardless of
whether the $\mathds{Z}_2$ subgroup or the $\mathds{Z}_3$ subgroup of the
gauge fields is dynamic and we will also use Binder cumulants to study the
phase boundaries for the $\mathds{Z}(6, 3)$ model.  In this model, the
parafermion edge modes of degree \(p=6\) are transmuted into Majorana edge
modes.
 
To determine the phase boundaries of both models, we use the crossings of the
Binder cumulants of system sizes $(N_1, N_2) = (32, 64)$ with internal bond
dimension $D = 64$.  For the $\mathds{Z}(6, 2)$ model, we investigated the
crossing along lines of constant $\kappa$ for $\kappa/J > 0.2$, whereas for
$\kappa/J < 0.2$, we focused on lines of constant $h$ in order to to ensure
that the critical line is hit at approximately right angle.  We proceeded
similarly for the $\mathds{Z}(6, 3)$ model.  The resulting phase diagrams of
the $\mathds{Z}(6,3)$ and $\mathds{Z}(6, 3)$ gauged parafermion chains are
shown in Fig.~\ref{fig:phase_diagram}.  There is a topologically non-trivial
$\mathds{Z}_3$ phase for the $\mathds{Z}(6, 2)$ model and a topologically
non-trivial $\mathds{Z}_2$ (Majorana) phase for the $\mathds{Z}(6, 3)$ model
in an extended region of the phase diagram.  Notice that the range of
$\mathds{Z}_3$ criticality in coupling space encompasses a line rather than a
point as for the standard \(\Z_3\) parafermion chain.  This observation may
prove useful in connection to a recent blueprint for obtaining Fibonacci
anyons;\cite{mong2014} we will come back to this point in the outlook.

Finally, let us emphasize that our numerical findings indicate that the two
critical lines share a common endpoint marking the end of the critical phase
of the \(\Z_6\) parafermion chain.  Obviously, our results are not totally
conclusive due to the limited system sizes and the lack of detailed
finite-size extrapolations in that area of the phase diagram.  However, we can
offer a qualitative explanation of this numerical observation.  It is a
well-investigated fact that the perturbed, ungauged parafermion chain (a.k.a,
the clock model) displays a critical phase for \(p>5\), see for example
Ref.~\onlinecite{ortiz:12} for a recent, comprehensive discussion.  Imagine
now for the sake of the argument that the transition line for the
$\mathds{Z}(6,2)$ model were to meet the critical phase of the \(\Z_6\)
parafermion chain somewhere inside that phase.  Then, from that point on, that
is, for larger values of \(h\), the chain would effectively behave as a
\(\Z_2\) chain, since the only edge modes left would be the critical Majorana
modes.  But a \(\Z_2\) chain does not support a critical phase.  Hence it must
be that the critical lines meet at the end of the critical phase.
Incidentally, this determines completely the boundary of the critical phase,
since the starting point is a function of the final point by self-duality.

\section{Summary and Outlook}
\label{outlook}

Topological zero-energy boundary modes are typically stable against
generic perturbations, and it is precisely this feature that makes them
attractive for quantum information processing.  However, it is also this
feature that makes them hard to mold in a controlled fashion.  Extending our
previous work on the effect of phase slips in the Majorana chain of Kitaev,
we have argued in this paper that for topological boundary modes associated 
to a global protecting symmetry, it is possible to modify the topological 
edge structure in a controlled fashion by engineering a gauge field into 
the system. The quantum fluctuations of the gauge field act as a relevant 
perturbation without symmetry breaking, and permit to split the ground 
degeneracy of the system, partially or completely, depending on the designed
properties of the gauge fluctuations. If the gauge field is engineered,
by allowing only restricted gauge fluctuations, so that the
ground degeneracy is only partially split, then the topological edge 
modes are modified accordingly, in a controlled and predictable fashion.  
There is absolutely no risk of driving the system to a trivial phase purely 
by restricted gauge fluctuations, unless the gauge field is explicitly 
designed to do so by allowing for unrestricted gauge fluctuations. It 
would be interesting to recast these results in terms of the group 
cohomology classification of symmetry protected topological quantum 
orders and bosonic anomalies.\cite{chen13,wang15}

We have illustrated our ideas with gauged parafermion chains.  For generic
chains, we performed a symmetry and duality analysis confirming our
theoretical predictions. To obtain a more refined picture of the 
the transmutation of the edge modes and the correlation with critical 
phenomena, we investigated numerically the \(\Z_6\) parafermion chain 
coupled to a \(\Z_2\)-like or a \(\Z_3\)-like gauge field. That is, the dynamics
of the physically natural \(\Z_6\) gauge field was restricted to achieve a 
\(\Z_2\)-like or \(\Z_3\)-like effect. According to our general picture, 
the \(\Z_2\)-like field should transmute the \(\Z_6\) edge modes of the 
parafermion chain into \(\Z_3\) edge modes, and we showed numerically that 
this is the case by computing the period of the supercurrent in 
a parafermion ring junction as a function of gauge fluctuations.
We also obtained the phase diagram of the gauged parafermion
chain as a function of gauge fluctuations and a perturbation driving the
system to a trivial phase. In this two dimensional phase diagram, the
transition between the gauge-driven phase with \(\Z_3\) edge modes and the
topologically trivial phase occurs on a critical line in the universality
class of the \(\Z_3\) parafermion chain. From this point of view, the
\(\Z_2\)-gauged \(\Z_6\) parafermion chain is indistinguishable from the
ungauged \(\Z_3\) chain.  The general picture is the same for the
\(\Z_3\)-gauged \(\Z_6\) parafermion chain. In this case, the edge modes 
are Majoranas.

Let us mention in closing two potential practical applications of our work. 
One possibility is to exploit our ideas to create topological qutrits out of
\(\Z_6\) parafermions. Qutrits have practical advantages over qubits,
not because they can provide qualitatively faster algorithms, but because 
they can polynomialy shorten the length of an algorithm in the circuit model of
quantum computation. Of course, the same holds for six-state quantum bits,
but here we have face the problem that very little is known about 
quantum software design with many-leveled logic. Topological qutrits strike 
a nice balance from this point of view. 

Another possibility is to use a gauged \(\Z_6\) parafermion chain as a 
replacement for a \(\Z_3\) chain. The \(\Z_3\) parafermion chains 
is one of the key building blocks in a recent blueprint\cite{mong2014} for realizing 
Fibonacci anyons. One difficulty of that blueprint is that, as a building 
block in this setup, all the chains in a two-dimensional stack must be 
tuned to their critical point. By contrast, our realization of the \(\Z_3\) 
chain out of the \(\Z_6\) chain is critical on a line rather than a point, 
and the whole line is in the universality class of the  \(\Z_3\) chain. 
Moreover, the \(\Z_6\) chain is in principle realizable out of the (doubled) 
\(1/3\) fractional quantum Hall state, which is the most stable fractional
plateau. We expect that the \(\Z_2\) gauge field for this application
might also be engineered out of superconducting phase slips, but the
situation is not as simple as for the Majorana chain. The mesoscopic details
of engineering the required gauge fields will be the subject of an upcoming
publication.

\acknowledgments

We gratefully acknowledge discussions with Michele Burrello, 
Kasper Duivenvoorden, Bernard van Heck, Jaakko Nissinen, Norbert Schuch,
and Dirk Schuricht. This work is part of the DITP consortium, a program 
of the Netherlands Organisation for ScientificResearch (NWO) that is 
funded by the Dutch Ministry of Education, Culture and Science (OCW).


\begin{thebibliography}{00}
\bibitem{footnote2}
One may think about the reintroduction of the gauge fields $\tau _j^z$ as the
  $\protect \mathds {Z}_2$ equivalent of the Peierls substitution used in
  lattice models to accomodate the effect of magnetic fields.
\bibitem{footnote3}
The phase diagram of the parafermion chain is isomorphic to that of the clock 
model, see Ref. \cite{ortiz:12} for an up-to-date discussion.
\bibitem{footnote4}
The reason for this requirement is the elementary relatioin
\(e^{\im2\pi/r}e^{\im 2\pi/s}=e^{i2\pi n(r+s)/rs}\). Recall that 
\(p=rs\).
\bibitem{footnote5}
The phase obtained from exchanging these two matrices is
\(e^{\im2\pi ns/r}=e^{\im2\pi sn(s+r)/p}=e^{\im2\pi s/p}=e^{\im2\pi/r}\).
\end{thebibliography}
\end{document}